\documentclass[aps,pra,10pt,notitlepage,twocolumn,superscriptaddress,nofootinbib,floatfix]{revtex4-2}
\pdfoutput=1
\usepackage[utf8]{inputenc}
\usepackage[T1]{fontenc}
\usepackage[english]{babel}
\usepackage{graphicx} 
\usepackage{xcolor}
\usepackage[bookmarks=false,colorlinks=true]{hyperref}

\begin{document}

\title{The path towards measuring the gravitational field of proton bunches at accelerators}

\author{Daniel Braun}
\affiliation{Institute for Theoretical Physics, Eberhard Karls Universität Tübingen, Auf der Morgenstelle 14, 72076 Tübingen}

\author{Rongrong Cai}
\affiliation{CERN - 1211 Geneva 23 - Switzerland}

\author{Pascal Hermes}
\affiliation{CERN - 1211 Geneva 23 - Switzerland}

\author{Marta Maria Marchese}
\affiliation{Institute of Science and Technology Austria (ISTA), 3400 Klosterneuburg, Austria}

\author{Stefan Nimmrichter}
\affiliation{Naturwissenschaftlich-Technische Fakult{\"a}t, Universit{\"a}t Siegen, Siegen 57068, Germany}

\author{Christian Pfeifer}
\affiliation{ZARM, Unversität Bremen, Am Fallturm 2, 28359 Bremen, Germany} 
\affiliation{Faculty of Mathematics and Computer Science, Transilvania University, Iuliu Maniu Str. 50, 500091 Brasov, Romania}

\author{Dennis R\"atzel}
\affiliation{ZARM, Unversität Bremen, Am Fallturm 2, 28359 Bremen, Germany}

\author{Stefano Redaelli}
\affiliation{CERN - 1211 Geneva 23 - Switzerland}

\author{Hendrik Ulbricht}
\affiliation{School of Physics and Astronomy, University of Southampton, SO17 1BJ, Southampton, UK.}

\begin{abstract}
The Newtonian law describing the gravitational interaction of non-relativistic (slowly moving) gravitating matter, has been tested in many laboratory experiments with very high precision. In contrast, the post Minkowskian predictions for the gravitational field of ultra-relativistic matter, dominated by momentum instead of rest mass, have not been tested directly yet. The intense ultra-relativistic proton beam in the LHC storage ring offers the potential to test general relativity and alternative gravitational theories in this parameter regime for the first time in controlled lab-scale experiments.  If successful, this would open the road to a novel use case of the LHC, where non-trivial gravitational physics could be studied likely in a parasitic mode, without the necessity of dedicated filling patterns.  While the technical challenges are formidable, they should also lead to the development of ultra-high-sensitive acceleration sensors with abundant applications in other parts of science and technology.  The present document summarizes the status of the theoretical studies in this direction, points out the challenges, and possible ways of addressing them.  It was submitted as a contribution to the European Strategy for Particle Physics (ESPP) 2026 Update.
\end{abstract}

\maketitle

\section{Scientific context}

Our understanding of the physics, from the smallest to the largest scales, is based on the geometry of spacetime as predicted by general relativity (GR) and the quantum field theories of the standard model of particle physics (SM). Since this understanding is incomplete, in particular in regard of the gravitational interaction, it is of utmost importance to map out the properties of gravity in all regimes to highest precision and compare them to theoretical frameworks that go beyond GR and the SM. A regime, that is underrepresented in highest precision studies of the gravitational field is the weak gravitational field of ultra-relativistic sources.

\subsection{The limits of general relativity and the standard model or particle physics and ways beyond}

The standard model of particle physics and the $\Lambda CDM$ model of cosmology based on General Relativity (GR) are the most successful description of nature from very small to very large scales that we have. 

Despite their numerous successes such as the prediction of gravitational waves, black holes, the motion of planets and stars, as well as the evolution of the universe, it is evident that they cannot be the final answer to our understanding of the gravitational interaction. Their incompleteness is easily demonstrated by puzzling observations and theoretical obstacles, such as the accelerated expansion of the universe, the rotation curves of galaxies, the Hubble ($H_0$) tension and the density ($S_8$ or $\sigma_8$) fluctuation tension in cosmology \cite{Abdalla:2022yfr}, the need for dark matter and dark energy (making up $\sim 27\% $ and $\sim 68\%$ of the universe \cite{Planck:2013oqw}), the prediction of singularities or infinite gravitational tidal forces, and the still elusive theory of quantum gravity (QG) \cite{Addazi:2021xuf}. Thus, clearly an extension of GR, the SM, or both, is needed for an improved understanding of gravity.

To improve our understanding of the gravitational force, numerous quantum and classical frameworks, models and theories have been suggested to extend GR, or the SM or both. An incomplete list of  such extensions include the Standard Model extension \cite{Kostelecky:2016uex}, Horndeski scalar-tensor gravity \cite{Kobayashi:2019hrl}, metric affine-, Poincar\'e gauge-, or teleparallel gravity \cite{JimenezCano:2021rlu,Hehl:1994ue,Obukhov:2022khx,Bahamonde:2021gfp} and Finsler gravity \cite{Heefer:2024kfi,Pfeifer:2019wus}. Up to date, none of them could explain satisfactorily all the aforementioned discrepancies between theory and observation or the theoretical difficulties in our description of gravity.

To scrutinize extensions of GR, the first applications and tests are usually performed in the context of cosmology~\cite{CANTATA:2021ktz}, by the study of compact objects \cite{Casado-Turrion:2023jba}, in a post-Newtonian weak field and slow velocity expansion \cite{damour1996tensor,Freire:relativ2012,will2014confrontation} and through their impact on gravitational waves signals \cite{Yunes:2025xwp}, meaning in highly symmetric situations, in the non-relativistic regime or from sources whose initial conditions are violent and subject to large uncertainties. 

More rarely, the impact of extensions of GR is investigated for very fast ultra-relativistic sources of gravity \cite{GERTSENSHTEIN}. One reason for this blank spot in the analysis of gravity theories is that the access to reliable measurements to falsify the theoretical predictions in this regime is sparse. One window to test the properties of the gravitational field of such sources are violent astrophysical environments, such as accretion discs or inspirals of binary systems shortly before they merge, where the ultra-relativistic matter leaves an imprint in gravitational wave signals \cite{Berti:2018cxi} or black hole observations. To identify these traces in observations is, however, very involved and subject to large uncertainties about the precise conditions of the source.

To remove the blank spot from the map of our understanding of gravity, we suggest to measure the gravitational field of ultra-relativistic matter sources to high precision in a highly controlled environment (compared to the astrophysical sources) that is available on Earth: the particle beams of particle accelerators.

The most controlled setup for the most relativistic particles on Earth are the proton bunches at the Large Hadron Collider (LHC). 
At the same time, the LHC also provides a circulating-beam line with the largest average power available, roughly 3.8$\cdot 10^{12}$\,Watt, orders of magnitude more than what is currently achievable with laser beams.  
The High-Luminosity upgrade of the LHC should increase the average power roughly by another factor two \cite{bruning_lhc_2004}.  
In addition, the LHC beam also offers the possibility of approaching the stored beam to a distance below a centimeter. Specially designed machine devices like beam collimators and roman pots for particle physics have proven the operation at distances as small as 1-2\,mm from the circulating beams, respectively.  While this needs to be demonstrated for a new experimental setup,  
sensitive detectors placed around the beam line of the LHC could be able to test the behavior of the gravitational 
near-field of this ultra-relativistic source 
of gravity, and the corresponding theoretical models, in  a very controlled environment \cite{Spengler:2021rlg}, complementary to the astrophysical observations.  
With the planned upgrade to the high-luminosity LHC and further technology development on the sensor side that appears feasible within a few years, this can lead to a whole new use case for the LHC.  Moreover, the project will lead to the development of ultra-sensitive acceleration sensors (see Section \ref{sec.readi}). These have tremendous significance in many areas of science and technology, from GPS-free navigation, over seismology, to geophysics and mineral exploration. 
In fundamental physics, the attempts of reaching experimentally the regime where both gravity and quantum mechanics play a role \cite{westphal_measurement_2020} will profit largely from their development. More long-term, one might even envisage creating quantum sources of gravity with measurable gravitational pull from the beam itself if the particle beam can be cooled close to the (transverse) ground state \cite{schroder_first_1990,hangst_laser_1991,sessler_cooling_1996,miesner_efficient_1996,litvinenko_coherent_2009,steck_cooling_2015,litvinenko_plasma-cascade_2018,krasny_high-luminosity_2020,brown_towards_2020} and squeezed or otherwise superposed in different quantum states. 

\subsection{The gravitational field of ultra-relativistic particle beams}
The gravitational field of an ultra-relativistic particle beam in particle accelerators, is mainly sourced by the momentum of the particles, while their rest mass only has a marginal contribution. Compared to the energy-momentum involved in astrophysical processes, the total energy-momentum of an ultra-relativistic particle beam is still small, which is why it can be analyzed in the weak gravity regime, however not in the non-relativistic post Newtonian regime. The mathematical framework that applies is the so called first order post-Minkowskian expansion. 

Due to the large circumference of the particle accelerators, compared to the test particle sensor, and the resulting short interaction time per bunch, 
it is justified to approximate the particle beam that sources the gravitational field with a source mass that passes the sensor moving on a straight line. 
The energy-momentum tensor of the source is proportional to the source masses' 4-velocity, and thus leads to an axially symmetric problem, where the direction of motion of the source defines the cylinder axis. Determining the gravitational field of this physical system is the gravitational analogue of the determination of the electromagnetic field of a relativistically moving charged particle that leads to the Li\'enard-Wiechert potential. 

The precise form of the gravitational field depends on the theory of gravity that is investigated. For general relativity and scalar-tensor theories, the precise form has been presented in \cite{Pfeifer:2024vvc}.

The reaction of the sensor to the gravitational field is predicted from the geodesic equation (or extensions thereof) which determines the acceleration of the sensor test particle in the gravitational field of the particle beam. The leading order contributions are given by the acceleration of the test particle transversal to the beam, that are caused by the gravitoelectric forces. Sub-leading transversal accelerations are caused by gravitomagnetic forces. 

The resulting momentum-transfer between the gravitational field of the particle beam and the test particle sensor is measurable and can be used to test predictions from general relativity and candidates for its extension.

\section{Objectives}
Measuring the gravitational field of ultra-relativistic particle beams to high precision opens the window to test general relativity and its extensions in a so far under-investigated parameter regime. 

Performing these measurements in a controlled environment on Earth, such as at particle accelerators, instead of deducing the gravitational field of ultra-relativistic sources from astrophysical cosmic observations which are subject to large uncertainties, is an important step towards an all-encompassing understanding of the gravitational interaction. Specifically, the objectives are:
\begin{itemize}
    \item To test GR in a new parameter regime, with an ultra-relativistic source of gravity in controlled lab-experiments,
    \item To test and possibly constrain alternative gravity theories,
    \item To develop ultrasensitive acceleration detectors,
    \item To develop distributed acceleration sensors.
\end{itemize}

\section{Readiness and expected challenges}\label{sec.readi}
The main challenge is the
extremely small signal. A cylindrical particle beam with average power $P$ leads to a gravitational near-field acceleration \cite{Spengler:2021rlg}
\begin{equation}
     a = \frac{4GP}{c^3\rho}
\end{equation}
at distance $\rho$, where $G$ is Newton's gravitational constant and $c$ the speed of light in vacuum.  For the proton beam at LHC, this boils down to an acceleration at 1\,mm distance from the beam center on the order $ a \sim 4\times 10^{-20}\,{\text{m}}/{\text{s}^2}$. This should be compared with the predicted reachable sensitivity of $\sim 10^{-14}$\,m/s${}^2/\sqrt{\text{Hz}}$ of accelerometers based on suspended magnets in the 10\,mm size-range \cite{PhysRevApplied.8.034002}. Three-axis electrostatic accelerometers with sensitivity of $\sim 3\cdot 10^{-12}$\,m/s${}^2/\sqrt{\text{Hz}}$ have  been demonstrated and are used on the ESA Gravity Field and Steady-State Ocean Circulation Explorer satellitte mission (GOCE), whereas the  acceleration sensors on the LISA pathfinder mission based on freely falling masses have achieved $\sim  10^{-14}$\,m/s${}^2/\sqrt{\text{Hz}}$ already. By measuring for about one week, the suspended magnets should reach a sensitivity of $\sim 10^{-17}$\,m/s${}^2$. Alternatively, in \cite{Spengler:2021rlg}, a monolithique pendulum was analyzed theoretically and it was shown that by cooling the relevant mechanical mode to nK temperature, optimizing its geometry and material, a signal-to-noise ratio of order 1 or even larger should be achievable with the high-luminosity LHC (see section \ref{sec.opt}). The expected signal-to-noise ratio is, however, only the theoretically achievable one taking into account thermal and quantum noise of the mechanical sensor.  Additional challenges arise from all other kinds of noise, such as read-out noise in continuous measurements,  seismic noise, electronic noise in the amplifiers, temperature fluctuations, and other technical noise.  In addition, the sensor will have to work in a radioactive environment and must be shielded from electromagnetic fields. These challenges will be addressed below.

\subsection{Status and challenges on the detector side }

The current state of art of quantum optomechanical acceleration sensing with pendulum-type probe masses is set by a high-Q milligram-scale monolithic pendulum \cite{catano-lopez_high-2020}. They demonstrated a mechanical Q-factor of about $2\times 10^6$ at a frequency $\omega_0=2\pi \times 2.2$\,Hz, which could be shifted with an optical spring to a range of frequencies from $2\pi\times$400\,Hz to $2\pi\times$1800\,Hz. This, however, then required additional electronic feedback cooling to compensate the heating through the optical spring and reduced the quality factor to about 250.  Comparing the expected sensitivity from the realized pendulum in \cite{catano-lopez_high-2020}, limited still by thermal noise (at an estimated temperature of a few mK), a gap in sensitivity after one week of measurement time of about 7 orders of magnitude still appears (see Table 2 in \cite{Spengler:2021rlg}), roughly on par with the  best existing acceleration sensors mentioned above.

\subsection{Theoretical detector optimization }\label{sec.opt}
  Based on the expression for the total expected noise power, both quantum and thermal,  the setup in \cite{catano-lopez_high-2020} was further optimized for maximum signal-to-noise ratio in the theoretical study \cite{Spengler:2021rlg}.  It was found that the S/N ratio increases with the square root of the product of quality-factor (including feedback cooling), mass and measurement time, and scales inversely with the resonance frequency.  This motivates the use of low-frequency sensors, where, however, the signal should have a matching frequency. Increase of the mass is limited by the implication of growing distance of the center-of-mass from the beam-line, but this constraint can be alleviated by a sensor geometry in the form of a long rod. Possibly this can be improved even further by having a pendulum body in the form of a half pipe, open to the beam, such that its center of mass can get closer to the beam line than its material components. With the optimization of a cylindrical rod made of silicon as in \cite{catano-lopez_high-2020} and allowed to be as long as 50\,cm, a mass of 33\,mg was found, if a distance of 200\,$\mu$m of the center of the detector from the beam axis was assumed.  Larger distances allow a quadratically increasing mass, but the reduced noise is to a large extent compensated by a decrease in the signal. Optimizing further over the remaining parameters (measurement time --- assumed up to 1 week), frequency, quality factor, and temperature of the relevant mode after feedback cooling, a signal-to-noise (S/N) ratio of about 0.6 was found in \cite{Spengler:2021rlg}. 
With the High-Luminosity LHC, and denser materials, such as tungsten, it was projected that this S/N ratio could be increased further to values of order 5-10. 
It remains to be seen whether technical constraints linked to accelerator operation, approaching the beam, requirements of shielding, and suppression of technical noises allow one to maintain a S/N ratio larger than 1.  

Alternative acceleration sensors based on levitated particles have been studied and developed in recent years, and experiments on optically, electrically, or magnetically trapping particles have achieved quality factors on the order of 10 million~\cite{vinante2020ultralow, westphal_measurement_2020, hofer2023high, fuchs2024measuring} for low frequency oscillators and acceleration sensitivity of $\sim 10^{-11}$\,m/s${}^2/\sqrt{\text{Hz}}$ have bee achieved by experiments~\cite{bose2025massive}. Levitation-based mechanical acceleration sensors provide a flexible platform that can be tuned across a wide range of frequencies - they are broadband - and, for example, by using metalens optical or cavity traps~\cite{sun2024tunable}, combs of detuned mechanical resonators can be implemented. As a general characteristic, acceleration sensitivity scales with the size of the probe mass~\cite{timberlake2019acceleration, rademacher2020quantum}, which can potentially be tuned to be very macroscopic even in the kilogram range~\cite{lecamwasam2020dynamics}. A particularly interesting platform for acceleration detection based on levitation is the diamagnetic levitation of graphite~\cite{tian2024feedback}, which has even been implemented for use in space and satellites~\cite{homans2025experimental}.
In addition, new ideas based on distributed sensing by using arrays of particles traps, i.e.~several or many mechanical sensors that get excited coherently due to the highly coherent gravitational source and read out coherently, have been studied; see Section \ref{sec.dis}.

\subsection{Signal-to-noise ratio increase through distributed sensing }\label{sec.dis}
Cascaded optomechanical systems are a promising approach to detect and accumulate the gravitational signal over multiple sensors across one or more locations around the LHC, with a potentially enhanced S/N ratio due to constructive interference.
By exploiting radiation-pressure interactions, optomechanical cavities facilitate an optical readout of mechanical signals, encoded in the quantum state $\rho$ of a mechanical resonator coupled to the optical cavity mode. By monitoring the cavity response to mechanical excitations, one can achieve extreme sensitivity to weak forces, including those from gravity; and sensors at different beam locations can reveal additional time information. However, practical implementations face the challenge of distinguishing this tiny signal from background noise, most notably thermal fluctuations and backaction noise, making precise measurement highly nontrivial.

When estimating a signal parameter $\theta$ from measurements on $\rho$, one usually gains precision by repeating the measurement $N$ independent times, e.g., on identically prepared probes. Averaging the noisy results improves the S/N ratio in proportion to $\sqrt{N}$, known as the Standard Quantum Limit (SQL). Quantum correlations, such as entanglement or squeezing, can further enhance the S/N ratio to scale as $N$, achieving the Heisenberg Limit (HL)~\cite{giovannetti2004quantum}. However, from an experimental perspective, establishing quantum correlations between many probes is highly challenging, as such quantum resources are particularly susceptible to decoherence. 
As a viable alternative, we propose to utilize the concept of coherent averaging~\cite{fraisse2015coherent} by cascading the optomechanical cavities along a single optical signal line. This line serves as a common quantum bus for the $N$ probes, along which the $N$ output signals interfere constructively. Instead of individual measurements, a single measurement is then performed on the bus to extract the coherently accumulated information about $\theta$, making HL scaling achievable in certain cases while remaining robust against decoherence, thus enhancing practical feasibility. This procedure is analogous to multipass schemes~\cite{higgins2007entanglement}, where HL is achieved without the need for entangled states, but simply by passing a single pulse multiple times through the same cavity.

Concretely, we assume that the $N$ initially uncorrelated optomechanical cavities are driven by a laser pulse sent through the quantum bus. 
The gravitational field signal is treated as an external impulse that sets the mechanical resonators into oscillation.
The coherent input light, modeled as a Gaussian pulse of width $1/\tau$, interacts with the mechanical resonators on a timescale much shorter than the mechanical period ($1/\tau\gg\omega_m$), while the cavity decay rate remains dominant ($\kappa\gg1/\tau$) in the unresolved sideband regime ($\kappa\gg\omega_m$). Our goal is to estimate the strength of the gravitational impulse from the phase and amplitude modulation of the detected output light. 

In the ideal case of lossless transmission and {phase-matched } oscillations, we find that the S/N ratio would reach HL scaling with $N$. More realistically, if we incorporate losses between the cascaded cavities along the quantum bus, the coherent gain in signal will be suppressed with growing $N$, leading to an optimal probe number $N_{\rm{opt}}$ beyond which the S/N ratio no longer improves and losses dominate.
Quantum noise due to thermal fluctuations and light backaction is expected to further decrease the S/N ratio. 
However, in the considered regime, the noise contribution from light backaction can be regarded as negligible, as it is quadratic in the optomechanical coupling. While thermal noise should increase amplitude fluctuations, 
we expect the coherent enhancement will not be significantly affected.

\subsection{Electromagnetic field of proton bunches and shielding }

Here, we briefly discuss the source's electromagnetic field and how to deal with it in a possible experiment. The employed sensors can be assumed to be nearly neutral, and therefore, electromagnetic forces arise mostly due to multipole moments and residual surface charges of the sensor masses. Still, these forces can be many orders of magnitude larger than the gravitational force on the sensor. We also need to take into account that the electromagnetic forces on the sensor mass oscillate at the same frequency as the gravitational force. The electromagnetic forces then drive a resonant sensor and the time dependence of the signal cannot be used to distinguish it from the expected gravitational signal. Hence, the sensor mass has to be shielded from the electromagnetic field of the source particles by a material layer with  appropriately chosen properties  (thickness, material, etc.). 

One could, for example, use a metallic shield which would exponentially suppress the time-averaged electric field on the scale of the Thomas-Fermi length which is roughly of the order of one \r{A} \cite{Ashcroft-Mermin}. Static magnetic fields can be shielded by superconducting materials on the length scale of the London penetration depth which ranges from few tens to hundreds of nanometers \cite{Kittel}. For the oscillating parts of the electric and magnetic fields, suppression is also exponential, however, on the scale of the much larger skin depth \cite{brandl2017towards} which is frequency dependent. In practice, damping of oscillating magnetic and electric fields by several hundred dB can be achieved \cite{brandl2017towards}.

To calculate the residual electromagnetic forces on the sensor mass, its composition has to be specified. Assuming the sensor mass to be composed of silicon dioxide and positioned at a distance of $1\,$cm from the beamline, we found that metal layers of a few millimeters should be sufficient to suppress the residual electromagnetic force, dielectrophoresis, below the size of the gravitational signal for the case of a beam that is modulated at a frequency of $100\,$Hz \cite{Pfeifer:2024vvc}. Furthermore, we found that this shielding should also be sufficient to suppress all magnetic effects and the Coulomb force due to remaining surface charges of the sensor mass.

\subsection{Impact of and shielding from stray particles }

One of the challenges from the accelerator point of view comes from the need to maintain an extremely low background level, such that gravitational 
momentum transfers per bunch
on the order of $\mathrm{10^{-31}}$~$\mathrm{kgms^{-1}}$ at distances of $5$-$10\,$mm could be identified. 

Beam losses occur naturally in all machine phases, including during beam collisions in all experiments. The LHC collimation system is designed to dispose safely and efficiently of these losses. However, the collimation efficiency is not perfect and depending on the location around the ring, different loss levels can be expected. 

Experimental areas are typically dominated by luminosity-debris losses. Therefore, in a first study, these areas were not favoured for physics experiments that need an installation close to the beams. A possible suitable location with relatively low losses and activations was identified in the Insertion Region 3 (IR3) of the LHC. 

To assess the potential impact of beam losses on the gravitational sensor, simulations with a preliminary experimental design has been carried out. First results suggest that we can expect a continuous momentum contribution from particle showers of one order of magnitude smaller than the gravitational signal. The contribution from primary particle loss should also not exceed 10\% of the gravitational signal. This preliminary study would then need to be completed for the final configurations in Run 4 by taking into account also luminosity losses emerging from the collision points.

In spite of an optimized design and positioning of the sensor, during one week of operation needed for averaging, it is likely to be hit by a high-energy particle. A primary proton from the particle beam could impart a  momentum to the sensor that is about 17 orders of magnitude larger than the expected transverse momentum transfer from the gravitational attraction of a single proton bunch. Such a relatively large momentum transfer could, however, be detected. One would then reset the sensor by cooling the relevant mode once more close to the ground state, which can be done on a time scale on the order of seconds, and restart a new measurement series, with the same timing relative to the filling pattern as the interrupted one.  While it remains to be examined in detail whether such numerical coherent signal addition can replace a long time series of measurements, such signal-processing and measurement strategy is also required due to the fact that fillings in the LHC have a lifetime of only several hours, much less than the averaging time of one week that was assumed in \cite{Spengler:2021rlg}.

\section{Timeline}

Fig.7 in ~\cite{bose2025massive} gives a summary of force noise versus probe mass achieved over the years from which one can deduct acceleration sensitivity.  For a given mass in the 100\,mg range one can see a trend of improvement of acceleration sensitivity by an order of magnitude over a typical time scale of a year or two, even though there is also an overall tendency to move to smaller masses. From this one might estimate that 3-5 more years might be needed to reach the required sensitivity of about $\sim 10^{-17}$\,m/s${}^2/\sqrt{\text{Hz}}$. 
At set of overall experimental milestones and estimated durations for achieving them is the following: MS1 setup levitated sensor experiment at underground lab in closed-cycle cryostat and with vibration isolation (2 years); MS2 Testing acceleration level in underground levitation experiment to push for $\sim 10^{-14}$\,m/s${}^2/\sqrt{\text{Hz}}$ (1 year); MS3 Further enhancement of sensitivity to $\sim 10^{-17}$\,m/s${}^2/\sqrt{\text{Hz}}$ (4 years);
MS4 Assembling levitated sensor next to LHC beam line (1 year).  Altogether this points at a timeline of about 8 years until the experiment can be operational. From the machine side alone, a simple installation could be done during an end-of-year technical stop if well prepared, so first  tests  could  already be envisaged in Run 4.

\section{Construction and operational costs}

In order to reduce acceleration noise to extreme levels below $\sim 10^{-13}$\,m/s${}^2/\sqrt{\text{Hz}}$, experiments have to be at low-noise settings. Thermal noise originating from the sensor environment is reduced by operating at low temperatures and is commonly achieved at a level of 10 mK in dilution cryostats, while it is possible to achieve lower temperatures of less than 1 mK with demagnetization stages which are now commercially available. Even lower temperatures are achieved for the sensor mode by cooling, for instance, for levitated mechanical systems now to the quantum ground state~\cite{magrini2021real}. This generates out of equilibrium situations which can be used to enhance sensitivities~\cite{jain2016direct}.

A sustainable way to cool to low temperatures in cryostats is by closed-cycle systems, which however have to be outfitted with advanced vibration isolation systems especially for low frequency sensors. This has been achieved recently~\cite{fuchs2024measuring} and LIGO-style geometric anti-spring (GAS) filters are a further option to achieve advanced vibration isolation (attenuation by three orders of magnitude of vibration amplitudes per filter stage) at low frequency of below 100 Hz. Closed-cycle cryostats also allow for very long detection run times of weeks and months for integration and averaging of statistical noises. Reduction of seismic noise is ideally done in underground laboratories, so we anticipate that a setup next to CERN-LHC is ideal for canceling seismic noises and by more than three orders of magnitude better than on the surface. 

Care has to be taken to not only reduce effects of mechanical vibrations and seismic noise, but also from electromagnetic noises, and specifically low-noise electronics has to be designed for the LHC probe experiment. For instance, levitated magnet-based sensors have to be shielded from earth and other external magnetic fields~\cite{ahrens2025levitated}. To reduce noise and decoherence effects of gas collisions, ultra-high vacuum technologies have to be combined with cryogenic technologies, which is possible, but not very common. Then extremely high vacuum of $\sim 10^{-17}$\,mbar at low temperature can be achieved~\cite{danielson2015plasma}. 

In summary, a dedicated experimental setup has to be designed to conduct the detection of gravity of the LHC beam. It should be located as close as possible to the particle beam and will be of table-top size. We estimate the total construction cost of such setup on the order of €5m. If a closed-cycle cryostat is used then there are no running costs beside those for electrical power. The same accelerometer for the detection of LHC beam gravity will also be a superb sensor for the hunt of Dark Matter candidates including Axion-like particles (ALPs)~\cite{amaral2024first, kalia2024ultralight, higgins2024maglev}.

\section*{Acknowledgments} This work was supported by the EU EIC Pathfinder project QuCoM (101046973). HU acknowledges financial support from UKRI EPSRC (EP/W007444/1, EP/V000624/1, EP/X009491/1) and STFC IAA Southampton grant, the Leverhulme Trust (RPG-2022-57), and the QuantERA II Programme (project LEMAQUME), Grant Agreement No 101017733. DR and CP acknowledge support by the excellence cluster QuantumFrontiers of the German Research Foundation (Deutsche Forschungsgemeinschaft, DFG) under Germany's Excellence Strategy -- EXC-2123 QuantumFrontiers -- 390837967. CP was funded by the Deutsche Forschungsgemeinschaft (DFG, German Research Foundation) - Project Number 420243324 - and by the Transilvania Felowships for Visiting Professors grant 2024 of the Transilvania University of Brasov. MMM acknowledges financial support though the project quantA, funded by  Austrian Science Fund (FWF) [10.55776/COE1].

\bibliography{refs}

\end{document}